\documentstyle[epsfig]{aipproc}
\begin{document}
\title{A Relic Neutrino Detector}

\author{C. Hagmann}
\address{Lawrence Livermore National Laboratory \\
7000 East Avenue, Livermore, CA 94550}

\maketitle

\begin{abstract}
Probably the most promising way of detecting cosmic neutrinos is measuring 
the mechanical force exerted by elastic scattering of cosmic neutrinos from macroscopic targets. 
The expected acceleration is $\sim 10^{-23} {\rm cm/s^2}$ for Dirac neutrinos of 
mass  $\sim 10\:{\rm eV}$ and local density $\sim 10^7/{\rm cm}^3$.
A novel torsion balance design 
is presented, which addresses the sensitivity-limiting factors of 
existing balances, such as seismic and thermal noise, and angular 
readout resolution and stability. 

\end{abstract}

\section*{Introduction}

The standard Big Bang model predicts a universal background of relic neutrinos, 
with an average density of $\sim 100/{\rm cm}^3$ per flavor. 
Relic neutrinos of mass $\gg 10^{-3}$~eV would be nonrelativistic today, and 
contribute to the cosmological energy density an amount 
$\Omega_\nu \sim \sum_{i} m_{\nu_i} /(90 h^2  {\rm eV})$, where
$h\simeq 0.65$ is the Hubble expansion rate 
in units of $100\:{\rm km/(s\, Mpc)}$.
Massive neutrinos are a natural 
candidate for the hot component in currently favored
``Mixed Hot+Cold Dark Matter (HCDM)'' \cite{primack98} 
models of galaxy formation. In this scenario, neutrinos would 
contribute $\sim$ 20~\%, and CDM (e.g. Wimps and axions) the remainder of the dark matter. 
Nonrelativistic neutrinos would 
be clustered around galaxies and move with a typical velocity  $v \sim 300\, {\rm km/s}$. 
The Pauli principle \cite{tremaine79} restricts the local neutrino number density to
$n_\nu \lesssim 2\times 10^6\rm cm^{-3}$ $(v_{\rm max}/10^{-3} c)^3$ 
$\sum_{i}(m_{\nu_i}/ {10 \rm eV})^3$.
The detection of relic neutrinos is hindered by the extremely small cross sections
and energy deposits expected from interactions with electrons and nucleons.
Past proposals have focused on detecting the mechanical force on macroscopic targets
due to the ``neutrino wind''.
Here, spatial coherence increases the cross section of targets smaller than
the neutrino wavelength $\lambda_\nu \sim 100 \,\mu {\rm m} (10 {\rm\, eV}/m_\nu)$.
In the nonrelativistic limit, one must distinguish between Majorana and Dirac 
neutrinos. For Dirac $\mu$ or $\tau$ neutrinos,
the cross section is dominated by the vector
neutral current contribution
$\sigma_ D=  (G_F^2 m_\nu^2/8\pi) N_n^2 =$$ 2\times 10^{-55} {\rm cm^2}
({m_\nu}/ {10 \rm \,eV})^2 N_n^2$,
where $N_n$ is the number of neutrons in the target
of size $\lesssim \lambda_\nu/2\pi $. For Majorana neutrinos, the 
vector contribution is suppressed
by a factor $(v/c)^2$.
The Sun's peculiar motion through the galactic halo will produce
a wind force, whose direction is modulated by the Earth's rotation.
For Dirac neutrinos, the acceleration of a target of 
density $\rho$ and radius $ \lambda_\nu/2\pi $ is \cite{shvarts82}
$a= 8\times 10^{-24} {\rm cm/s^2} $$((A-Z)/A)^2 $
$(v_{\rm sun}/10^{-3}c)^2$
$(n_\nu/10^7{\rm cm^{-3}})$
$(\rho/20\,{\rm gcm^{-3}})$
and is independent of $m_\nu$. 
A harmonic oscillator driven by the
neutrino wind on resonance 
would experience a displacement amplitude of 
$\Delta x\simeq 10^{-15}{\rm cm}(\tau/{\rm day})$. 
A target of size $\gg\lambda_\nu $ can be assembled, 
while avoiding destructive interference, by using foam-like
or laminated materials \cite{shvarts82}.
Alternatively, grains of size  $ \sim \lambda_\nu $ could be randomly embedded
in a low density host material.

\section*{Proposed Detector}
At present, the most sensitive detector of small forces is the 
``Cavendish'' torsion balance, 
which has been widely used for measurements of $G$, 
searches for new forces, and tests of the equivalence principle. 
A typical arrangement consists of a  
dumbbell-shaped test mass suspended by a 
tungsten fiber.
The angular deflection is read out with
an optical lever. 
The most serious noise backgrounds are 1. thermal noise, 
2. seismic noise, 3. time-varying
gravity gradients.
The smallest measurable acceleration is $\sim 10^{-12} \rm cm/s^2$
\cite{adel91}.   
Several improvements seem possible:
Thermal noise can be decreased by lowering the temperature and by
employing a low dissipation (high-$Q$) suspension,
as seen from the expression for the thermal noise acceleration\cite{brag92book}  
$a_{\rm th} = 2\times 10^{-23} {\rm {cm}/{s^{2}}}\;$ $(T/{\rm K})^{1/2}$
                               $({\rm 1 day}/{\tau_0})^{1/2}$
                               $({10^6 \rm s}/{\tau})^{1/2}$
                               $({10^{16}}/{Q})^{1/2}$,
where $\tau$ is the measurement time, $\tau_0$ is the oscillator period, 
 and $T$ is the operating temperature.
A promising low-dissipation
suspension method uses the Meissner effect. 
Niobium or NbTi based suspensions have been employed in gravimeters, gyros, 
and gravitational wave antennas.
Generally, the magnetic field applied to the superconductor is limited
to $\lesssim 0.2\rm T$ to avoid flux penetration or loss of
superconductivity. 
A remaining problem is  
flux creep noise and dissipation because of incomplete flux expulsion.
An alternative method would employ a passive
persistent-mode superconducting magnet floating above
a fixed suspension magnet as shown in Fig. 1. This allows
a much higher lifting force because the critical field of NbTi wire
is several T.
Moreover, the flux lines are strongly pinned by artificial wire defects, 
leading to a small field decay rate of
$\dot B/B \sim 10^{-8}/\rm hour$.
The cylindrical symmetry of the suspension magnets
allows a very long rotational oscillation period,
which can be matched to the diurnally varying neutrino wind
by applying a suitable restoring force.
The effects of seismic noise and gravity gradients can be
reduced with a highly symmetric target (see Fig. 1).   
With the c.m. of the target centered below the suspension support, 
the leading order gravitational torques arise from the dipole and
quadrupole moments of the target which need to be minimized
by balancing.
Braginsky et.al. \cite{brag77} have given estimates of the seismic power
spectra for vertical and horizontal, and rotational seismic modes.  
For example, the horizontal acceleration at
$\omega\sim 10^{-4}\rm s^{-1}$ is
$a\sim 10^{-12}{\rm cm/s^2}$$(\Lambda/100 {\rm \,km})$$(10^6/\tau)^{-1/2}$, where
$\Lambda$ is the seismic wavelength. 
Hence, the coupling of this mode to the wanted rotational mode of the torsion
oscillator must be made
very small. Especially worrisome is rotational seismic noise 
which will directly mask the signal and needs to be
compensated. 
\begin{figure}[t] 
\centerline{\epsfig{file=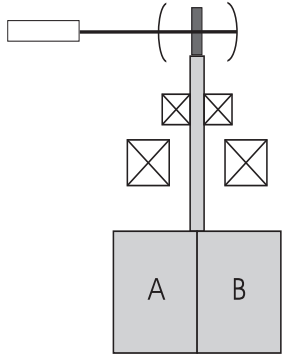,height=2.0in,width=2.0in}}
\vspace{10pt}
\caption{Schematic diagram of the torsion oscillator. The target consists
of two hemicylindrical masses with similar densities but different
neutrino cross sections. The mass is suspended by a ``magnetic hook''
consisting of a superconducting magnet in persistent mode
floating above a stationary magnet. The rotation angle is read out by
measuring the resonance frequency of a tunable Fabry-Perot cavity
with a laser.} 
\label{fig1}
\end{figure}
The proposed angular rotation readout
is composed of a parametric transducer, which converts the angle to
an optical frequency. As shown in Fig. 2, the transducer
consists of a high-$Q$ optical cavity of length $l$, tuned by a Brewster-angled low loss
dielectric plate of thickness $d$. A cavity finesse $F\simeq 10^5$ 
should be obtainable with dielectric mirrors  
for the Gaussian ${\rm TEM}_{00p}$ modes.
The frequency tuning sensitivity is $\Delta f/f \sim (d/l)\Delta\theta$ 
yielding a resolution of $\sim 10^{-14} \rm rad/Hz$. The angular measurement 
precision depends on the number of photons $N$ and laser wavelength $\lambda$ via 
$\Delta\theta\sim \lambda/(dF\sqrt{N})$. This is a factor $\simeq F$ better than
the optical lever for the same laser power. 
Cryogenic optical resonators have excellent long-term stability and have been
proposed as secondary frequency standards. The measured frequency drifts 
range from
$\sim 1\,\rm Hz$ over minutes to $\sim 100\,\rm Hz$ over days\cite{seel97}. 
For the measurement of the rotation angle, a stable reference frequency will be
required. This can be implemented using a laser locked to a second 
(untuned) cavity.  
Because of its symmetry, the described angle readout
 has high immunity against lateral, tilt, and vertical
seismic noise, but couples to rotational noise. A possible solution would be to 
suspend the target as well as the optical cavity in order to suppress the
common rotation mode.
Additional background forces will arise from gas collisions, cosmic ray hits, and 
radioactivity, etc., resulting in a Brownian motion of the target.  
The equivalent acceleration is $a\sim (\bar p/m)(\Gamma/\tau)^{1/2}$, where $\bar p$ is
the average momentum transfer, $\Gamma$ is the collision or decay rate, and
$m$ is the test mass. 
The residual gas pressure must hence be kept very low 
by cryopumping and the use of getters.
The cosmic muon flux
at sea level is $\simeq 100 \rm/(m^2s)$ and
can be reduced by going underground.
Further disturbing forces may be caused by
time-varying electric and magnetic background fields, which can
be shielded with superconducters. Thermal radiation and radiometric effects 
are much reduced by a temperature controlled cryogenic
environment. 
\begin{figure}[t] 
\centerline{\epsfig{file=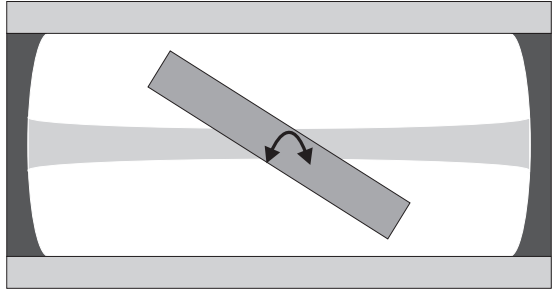,height=1.60in,width=3.0in}}
\vspace{10pt}
\caption{Topview diagram of the tunable high-$Q$ optical cavity.
The cavity is operated in the Gaussian $\rm TEM_{00p}$
mode.
 A narrow linewidth
$\sim \rm 100\,kHz$ is achieved with high reflectivity mirrors and
a low-loss dielectric Brewster-angled plate.}
\label{fig2}
\end{figure}
Finally, there is a fundamental limit imposed by the uncertainty 
principle, with a minimum measurable acceleration given by 
$a_{\rm SQL} = 5\times 10^{-24}{\rm cm/s^2}$$(10{\rm kg}/m)^{1/2}$
$({\rm 1 day}/\tau_0)^{1/2}$
$({10^6 \rm s}/\tau)$.
In our proposed position readout, the disturbing back action force arises from
spatial fluctuations in the photon flux passing through the central tuning plate.
There is hope that relic neutrinos will be detected in the laboratory early in
the next century, especially if they have masses in the eV range and are of 
Dirac type. The most viable way seems to be a much improved torsion balance
operating underground.
Naturally, a slightly modified balance could be used
to test the equivalence principle \cite{roll64} and to search for
new macroscopic forces \cite{adel91}.

\section*{Acknowledgements}
This research was performed under the auspices of the U.S. Department
of Energy by the Lawrence Livermore National Laboratory 
under contract W-7405-ENG-48.

\end{document}